\documentclass{article}
\pdfoutput=1

\usepackage{arxiv}

\usepackage[utf8]{inputenc} % allow utf-8 input
\usepackage[T1]{fontenc}    % use 8-bit T1 fonts
\usepackage{hyperref}       % hyperlinks
\usepackage{url}            % simple URL typesetting
\usepackage{booktabs}       % professional-quality tables
\usepackage{amsfonts}       % blackboard math symbols
\usepackage{nicefrac}       % compact symbols for 1/2, etc.
\usepackage{microtype}      % microtypography
\usepackage{graphicx}
\usepackage{setspace}
\usepackage{amsmath}
\usepackage{subfigure}
\usepackage{multirow}
\usepackage{algorithm, algorithmic}
\usepackage{cleveref}

\title{Neural network with data augmentation in multi-objective prediction of multi-stage pump}

\author{
  Hang Zhao \\
  Department of mechanical engineering\\
  Zhejiang University\\
  N0.38 Zheda Road, Hangzhou, China \\
  \texttt{zhaoyihang@zju.edu.cn} \\
}

\begin{document}
\maketitle

\begin{abstract}
A multi-objective prediction method of multi-stage pump method based on neural network with data augmentation is proposed. First of all, the hydraulic performance predictions of a multi-stage centrifugal pump based on surrogate models and neural network are investigated. Then the sensitivity analysis of design variables is implemented and three key design variables are determined to be optimized base on hydraulic loss model. In order to study the highly nonlinear relationship between key design variables and centrifugal pump external characteristic values (head and power), the neural network model (NN) is built in comparison with the quadratic response surface model (RSF), the radial basis Gaussian response surface model (RBF), and the Kriging model (KRG). The numerical model validation experiment of another type of single stage centrifugal pump showed that numerical model based on CFD is quite accurate and fair.  All of prediction models are trained by 60 samples under the different combination of three key variables in design range respectively. The accuracy of the head and power  based on the four predictions models are analyzed comparing with the CFD simulation values. The results show that the neural network model has better performance in all external characteristic values comparing with other three surrogate models. Finally, a neural network model based on data augmentation (NNDA) is proposed for the reason that simulation cost is too high and data is scarce in mechanical simulation field especially in CFD problems. The model with data augmentation can triple the data by interpolation at each sample point of different attributes. It shows that the performance of neural network model with data augmentation is better than former neural network model. Therefore, the prediction ability of NN is enhanced without more simulation costs. Neural network is a powerful and promising tool to fit the highly nonlinear relationship between the key design variables and the pump external characteristic values. With data augmentation it can be a better prediction model used in solving the optimization problems of multistage pump for next optimization  and generalized to finite element analysis optimization problems in future.
\end{abstract}

% keywords can be removed

\section{Introduction}
\label{intro}
Centrifugal pumps are broadly utilized in mining drainage, urban water supply and drainage, farmland water system, petrochemical and nuclear power system, representing about 70\% of the all kinds of pumps and the average yearly power utilization accounts for above 10\%. Subsequently, optimization of the centrifugal pump isn't just of extraordinary importance to improve the performance of centrifugal pumps, yet in addition create gigantic social and economic benefits. there is a great number of multistage pump structure optimizations based on hydraulic loss model. For example, hydraulic loss model is used to optimize the outlet angle of leaves, the quantity of leaves to improve the efficiency of centrifugal pump under the design condition by Yang Junhu \cite{Yang2016}. Nie Songhui et al. took the hydraulic loss model as the optimization calculation model, used the theoretical disparity of no hump and no overload as the limitation condition to improve the hydraulic efficiency of the centrifugal pump \cite{Nie2013}. Abdellah Kara Omarand et al. built up a hypothetical program to foresee the performance of centrifugal pumps utilizin both theoretical and empirical internal and external energy loss equations. The head, power and efficiency of centrifugal pumps with various geometrical shapes, speeds and fluid properties (viscosities) are predicted. Good agreement between prediction and test results was achieved \cite{KaraOmar2017}.

With the development of computer fluid dynamics and computer parallel computing ability, noteworthy researches are focused on utilizing the surrogate models based on computer fluid dynamics (CFD) to improve the performance of different types of pumps. Wang Chunlin et al. built up the quadratic response surface surrogate model for the propose of optimizing the efficiency, the width of the high efficiency zone and the performance of the no hump \cite{Wang2013}. In order to optimize the blade inlet angle, angle of wrap and outlet angle under two working conditions, Wang Wenjie et al implemented the Kriging model to improve the efficiency of centrifugal pump \cite{Wang2015}. The utilization of surrogate models is proposed by Sayed Ahmed Imran Bellary for electrical submersible and industrial centrifugal pumps to optimize the inlet and outlet angles\cite{Bellary2014, Bellary2016, Bellary2016a}. Man-Woong Heo et al. optimized the quantity of blades, the shape of the hub and shroud, the inlet and outlet angles of the blade by means of three surrogate models \cite{Heo2016}. L Zhou et al. examined the impact of impeller rear shroud radius to the axial force and pump hydraulic performance \cite{Zhou2013}. A Fleder and M Böhle studied two modular side channel pump models both numerically and experimentally. For both particular designs, diverse side channels and impellers could be studied, with the goal to get data about the influence of the various geometries on the performance, and the inner flow phenomena of the pump \cite{Fleder2015}.  Mortazavi et al. examined the influence of blade angle on the performance of a regenerative pump \cite{Mortazavi2017}. A Patil et al. give pump users a straightforward way to foresee a pump's performance change under the change of fluid viscosity \cite{Patil2019}. DDO Weme et al. used prediction methods from literature to quantify the effect of trimming on the hydraulic performance, together with a new prediction method based on the simplified description of the flow field in the impeller \cite{Weme2018}. Tong et al studied the Multi-objective optimization of multistage pump based on surrogate model \cite{shuiguang2020}.

The key structure variables chosen in each literature are not the equivalent. There is no unmistakable and perfect selection method. If too many design variables are selected, at that point the establishment of the surrogate models demands an enormous number of samples, devouring huge time consume and resources. Nonetheless, if the key design variables are sensibly selected, the surrogate models can be utilized to assess the performance of the centrifugal pump including the highest efficiency, the width of the high effectiveness zone and the cavitation erosion resistance with sensible and adequate samples. The above two methods and flow field calculation method are mostly used, Although flow field calculation method and hydraulic loss method with the performance prediction accuracy of centrifugal pump are compared \cite{Tan2005}, hydraulic loss method and surrogate model are all carried out separately without correlation of estimation precision and adequacy.

\begin{figure}[tbh]
	\centering
	\includegraphics[scale=1.2]{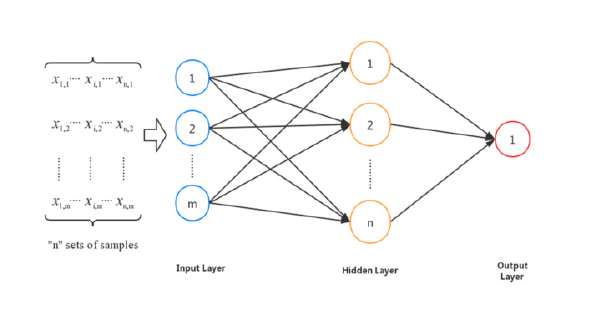}
	\caption{Structure of a neural network.}
	\label{fig NNstructure}
\end{figure}

A neural network is an artificial neural network, composed of artificial neurons or nodes, which is becoming a vital tool for solving AI problems, such as pattern recognition (PR), natural language processing (NLP). The structure of a simple neural network is shown in Fig.~\ref{fig NNstructure}. The starter hypothetical base for contemporary neural networks was independently proposed by Alexander Bain \cite{Bain1873mind} and William James \cite{James1890}. In their work, both body movement and thoughts resulted from collaborations among neurons inside the mind. The associations of the natural neuron are demonstrated as weights. A positive weight mirrors an excitatory association, while negative values mean inhibitory associations. All sources of info are modified by a weight and summed. At last, an actuation function controls the amplitude of the output.  This data-driven methodology is reasonable for many empirical data sets, wherein hypothetical directions is inaccessible to recommend an appropriate data generation process. The back propagation neural network (BPNN) is a standout amongst the most used forward neural networks \cite{Wang2006}. BPNN, otherwise called error back propagation network, is a multilayer-mapping network that minimizes an error backward while data is transmitted forward. It has shown that neural network is playing an important role in machine learning field and evolves rapidly. As a powerful tool, it's promising to apply neural network to optimization problem of multistage pump, which is not common in the now available literature.

In design and optimazation of mechanical field, the idea of data augmentation arises naturally in facing lack of value problems. One experiment or simulation may require too much effort and time. For example, CFD simulation is well-known for its time-consuming feature in engineering analysis When facing plenty of  combinations of design conditions, hence the simulation work becomes tedious excluding the technical operations. Thus data augmentation means to a scheme of augmenting the data that have been obtained so as to make it more convenient to analyze. This scheme is used to great advantage by the EM algorithm  in solving maximum likelihood problems \cite{dempster1977maximum}. Nowadays in machine learning problems it still needs large labeled dataset to succeed, and dataset is the vital part for training the satisfying models. However, the disadvantage in optimization problems of mechanical engineering is that the number of dataset we can get is commenly tens and hundreds, even less. Therefore, data augmentation is of great significance in mechanical optimazation problems, especially with CFD analysis.
%%%%%%%%%%%%%%%%%%%%%%%%%%%%%%%%%%%%%%%%%%
\section{ Key design variables determination  and sensitivity analysis}
\label{sec:1}
In this paper, the structure prerequisites of the 10-stage centrifugal pump are the flow (\emph{Q}) 100 $m^3/h$, the head (\emph{H}) 80 $m/stage$, the rotary speed (\emph{n}) 2950 $r/min$, and the rated output power (\emph{P}) 355 \emph{KW}. The specific speed is $n_s$  can be determined as
\begin{equation}
{n_s} = \frac{{3.65n\sqrt Q }}{{{H^{3/4}}}} = 67.09
\end{equation}

\begin{figure}[tbh]
	\centering
	\includegraphics[scale=1.2]{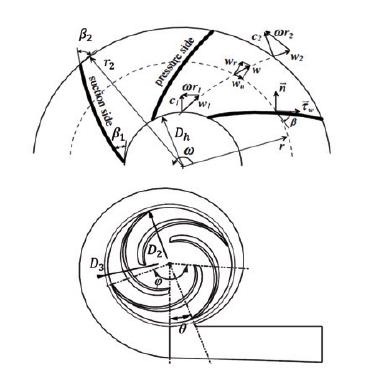}
	\caption{Impeller schematic and key design variables.}
	\label{fig Impeller}
\end{figure}

GB 19762, the national standard of China, \emph{Water centrifugal pump energy efficiency limit value and energy-saving rating value} point out that pump restricted efficiency is 72.9\% and pump target energy efficiency limit is 70.9\% for multi-stage water centrifugal pump under the design flow.
There are plenty of centrifugal pump design variables, including the number of blades $Z$, the inlet diameter $D_s$, $D_h$ of the hub and shroud, the blade inlet width $b_1$, the blade inlet angle $\beta_1$, the impeller outlet diameter $D_2$, the blade outlet width $b_2$, the blade outlet angle $\beta_2$, the blade warp angle $\phi$, blade thickness $S$, volute base diameter $D_3$, volute tongue angle $\theta$ and volute throat area ratio  $Y$. The impeller schematic is shown in Fig.~\ref{fig Impeller}.

Considering the high similarity of each pump stage and reduction of simulation cost, the first impeller is researched in this paper. As indicated by the customary speed coefficient strategy and the experience of designing a huge number of excellent practical pumps, the ranges of the design variables for the middle-opening multi-stage centrifugal pump in this paper are appropriately augmented and presented in Table~\ref{table1}.
\begin{table}[thp]\footnotesize
\setstretch{1.5}
\centering
\caption{Range of design variables} \label{table1}

\addtolength{\tabcolsep}{4.8pt}
\begin{tabular*}{7.95cm}{cc}
	\toprule[0.75pt]
     Design variables    &		  Range					 \\
     
	\midrule[0.5pt]
	   $  Z		   $     &    [3, 7]   						       \\
     $ \beta_1 $     &    $[10^{\circ},  35^{\circ}]$              \\
	   $ \beta_2 $     &    $[{14^ \circ }, {24^ \circ }]$           \\
	   $ D_2     $     &    $[10.4{\left( {\frac{{{n_s}}}{{100}}} \right)^{ - \frac{1}{2}}}\sqrt[3]{{\frac{Q}{n}}},  11.1{\left( {\frac{{{n_s}}}{{100}}} \right)^{ - \frac{1}{2}}}\sqrt[3]{{\frac{Q}{n}}}]$      \\
	   $ b_2	 $     &     $[0.85{\left( {\frac{{{n_s}}}{{100}}} \right)^{\frac{5}{6}}}\sqrt[3]{{\frac{Q}{n}}},  1.2{\left( {\frac{{{n_s}}}{{100}}} \right)^{\frac{5}{6}}}\sqrt[3]{{\frac{Q}{n}}}]$              \\
	   $ D_s	 $     &    $[12\sqrt[3]{{\frac{Q}{n}}},  15\sqrt[3]{{\frac{Q}{n}}}]$     \\
	   $ D_h     $     &   $[8.5\sqrt[3]{{\frac{Q}{n}}},  11.7\sqrt[3]{{\frac{Q}{n}}}]$   \\
       $ \phi 	 $	   &   $[{140^ \circ }, {180^ \circ }]$   \\
	   $ \theta  $	   &   $[{18^ \circ }, {28^ \circ }]$     \\
	   $ D_3	 $     &   $[1.03{D_2}, 1.06{D_2}]$          \\
	   $ Y		 $     &    $[0.8, 2]$                 \\     
	\bottomrule[0.75pt]
\end{tabular*}
\end{table}
According to the above design constraint range and hydraulic loss calculation model in  \cite{Tan2007}, the influence of a solitary design variable on the hydraulic performance is researched, that is to say, the design variables are decoupled. The sensitivity analysis of the design variables are implemented in the monotonous interval. The equation is written as follows:
\begin{equation}
\varepsilon {\rm{ = }}\frac{{{\rm{(}}{f_1} - {f_0})/{f_0}}}{{2\% }}
\end{equation}

\begin{figure}[tbh]
	\centering
	\includegraphics[scale=1]{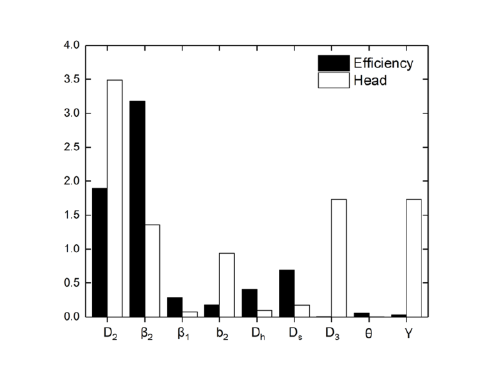}
	\caption{Sensitivity analysis of centrifugal pump design variables.}
	\label{fig sensitivity}
\end{figure}

Where $f_1$  is the efficiency or head after specific design variable increases by 2\%, $f_0$  is the efficiency or head when the design variables are all default values. The sensitivity analysis of design variables can be seen in Fig.~\ref{fig sensitivity}.

In the light of the outcomes of sensitivity analysis, it is realized that the integrated effect of impeller outlet diameter $D_2$, impeller outlet width $b_2$  and outlet angle $\beta_2$ is the most influenceable design variables. Therefore, the optimization of $x = \left[ {{D_2},{b_2},{\beta _2}} \right]$  for the hydraulic performance is of great significance.
\section{Constructions of surrogate models and neural network model} \label{sec:2}
By the quadratic response surface model (RSF), the radial basis Gaussian response surface model (RBF) and the Kriging model (KRG) under 0.7 Q, 1.0 Q and 1.2 Q working conditions, the highly nonlinear connections between the key design variables and the performance values are established. The root mean square error ($RMSE$) and coefficient of determination ($R^2$) are used to measure the validity test. The equations are as follows:
\begin{equation}
RMSE = \frac{1}{{m{{{\rm{\bar y}}}^{cfd}}}}\sqrt {\sum\limits_{i = 1}^m {{{({y_i}^{cfd}{\rm{ - y}}_i^{rsf})}^2}} }
\end{equation}

\begin{equation}
{R^2} = 1 - \frac{{\sum\limits_{i = 1}^m {{{({y_i}^{cfd}{\rm{ - y}}_i^{rsf})}^2}} }}{{\sum\limits_{i = 1}^m {{{({{\rm{y}}_i}^{cfd} - {{{\rm{\bar y}}}^{cfd}})}^2}} }}
\end{equation}

Where  $m$ is the sample number of sample training set,  ${y_i}^{cfd}$ is the numerical calculation, ${\rm{y}}_i^{rsf}$ is response surface model calculation, and ${y_i}^{cfd}$  is the average value of numerical calculation. When the root mean square error closer to 0 and the coefficient closer to 1, the fitting accuracy is higher.
Considering the dimension of the key design variables, the RSF is divided into a complete quadratic response surface model with 10 unknown parameters. The response functions are
\begin{equation}
  \begin{aligned}
    {y} & = {a_0} + \sum\limits_{i = 1}^3 {{a_i}{x_i}}  + \sum\limits_{j = 1}^3 {{a_j}} {x_j}^2 + \sum\limits_{i = 1}^3 {\sum\limits_{j = 1,j \ne i}^3 {{a_{ij}}} {x_i}{x_j}} \\
     & = \left[ {1,x,{x^2},{x_1}{x_2},{x_1}{x_3},{x_2}{x_3}} \right] \cdot C
  \end{aligned}
\end{equation}

The regression coefficients $a_0$, $a_i$, $a_j$, $a_ij$ respectively are the constant term, the primary term, the quadratic term, the cross terms of the RSF model. $C$ is the coefficient matrix of the quadratic response surface function.

The radial basis Gaussian response surface model is generally divided into the input layer, hidden layer and output layer. The hidden layer neuron transform function, to be specific as radial basis core function using, uses Gaussian function as follows.
\begin{equation}
  d\left( r \right) = {e^{ - {r^2}}} = \exp \left( { - \frac{{{{\left\| {X - {X_i}} \right\|}^2}}}{{2{\sigma ^2}}}} \right)
\end{equation}
The output of the radial basis Gaussian response surface is
\begin{equation}
{y}\left( {x} \right) = \sum\limits_{i = 1}^n {{w_i}} \exp \left( { - \frac{{{{\left\| {X - {x_i}} \right\|}^2}}}{{2{\sigma ^2}}}} \right)
\end{equation}

Where  $X_i$ is the center of Gaussian function, $w_i$  is the weight coefficient, and $X$ is all the samples in the training set. The above equation can be converted to the following matrix form:
\begin{equation}
  {y} = {D_x} \cdot w
\end{equation}

The  $y$ as 60 by 1 output matrix is utilized to describe the centrifugal pump performance, $D_x$  is a 60 by n dimensional Gaussian distance matrix,  $w$ is n by 1 dimensional weighting factor matrix which is obtained through training the 60 samples using RBF. Since the example training set has 60 samples, we generally choose n = 10 or n = 20 or n = 30. Comparing the root mean squared error and coefficient of determination, we choose n = 30. So the optimization model acquired by RBF is:
\begin{equation}
  f\left( x \right) = D \cdot w
\end{equation}

Where $D$ is the 1 by n Gaussian distance matrix of the key design variables $x$,  $f\left( x \right)$ is the response function.

The Kriging model has global characteristics, which is utilized to predict the value of centrifugal pump performance based on 60 sample pumps. The expression is made of regression model and Gaussian correlation model in two parts:
\begin{equation}
  {y}({x}) = \beta f({x}) + z({x})
\end{equation}

Where  $y(x)$ is the response function, $f(x)$  is the global basis function, $\beta$  is the regression coefficient, and $z(x)$ is the Gaussian correlation function whose mathematical expectation and covariance are:
\begin{equation}
  \left. \begin{array}{l}
E(z(x)) = 0\\
Cov(z({x_i}),z({x_j})) = {\sigma ^2}R(\theta ,{x_i},{x_j})\\
R(\theta ,{x_i},{x_j}) = \prod\limits_{d = 1}^q {\exp ( - {\theta ^d}{{\left| {x_i^d - x_j^d} \right|}^{{p^d}}})}
\end{array} \right\}
\end{equation}

Where  ${\sigma ^2}$ is the variance,  $\theta $ is unknown association parameter, $R(\theta ,{x_i},{x_j})$  is the correlation function between point ${x_i}$ and ${x_j}$ . The predictive value of the point $x$  of the Kriging substitution model is:
\begin{equation}
  \mathop y\limits^ \wedge  (x) = \mathop \beta \limits^ \wedge  f(x) + {r^T}(x){R^{ - 1}}(Y - \mathop \beta \limits^ \wedge  F)
\end{equation}

Where ${r^T}(x)$  is the vector of relationship between the unknown point and the known point.

A single hidden layer BPNN comprises of an input layer, a hidden layer, and an output layer as shown in Fig.~\ref{fig NNstructure}. Adjacent layers are associated by weights, which are constantly distributed between -1 and 1. A perfect theory to decide the number of input nodes and hidden layer nodes is unavailable. A single hidden layer BPNN is utilized for onestep-ahead forecasting. A few past observations are utilized to estimate the present value. The equations are shown as follows:
\begin{equation}
  {I_j} = \sum\limits_{i = t - n}^{t - 1} {{w_{ji}}}  \times {y_i} + {\beta _j}\quad (j = 1, \ldots ,h)
\end{equation}
\begin{equation}
  {y_j} = {f_h}\left( {{I_j}} \right)\quad (j = 1, \ldots ,h)
\end{equation}
\begin{equation}
  {I_o} = \sum\limits_{j = 1}^h {{w_{oj}}}  \times {y_j} + {\alpha _o}\quad (o = 1)
\end{equation}
\begin{equation}
  {y_t} = {f_o}\left( {{I_0}} \right)\quad (o = 1)
\end{equation}

Where $I$ means the input, $y$ signifies the output,  $y_t$ is the predicted value of point $t$, $n$ and $h$ are the numbers of input layer nodes and hidden layer nodes, respectively. ${w_{ji}}$  is the connection weights of the input and hidden layers. ${w_{oj}}$  is the connection weights of the hidden and output layers.  ${\beta _j}$ and ${\alpha _o}$  means the threshold values of the hidden and output layers, separately, which are generally distributed between -1 and 1. ${f_h}$  and ${f_o}$  are the activation functions of the hidden and output layers, respectively. For the most part, the activation function of each node in the same layer is the same. The most broadly utilized activation function for the output layer is the linear function on the ground that the nonlinear activation function may introduce distortion to the estimated output. Researchers often uses the logistic and hyperbolic functions as the hidden layer activation functions \cite{Zhang1998}.

In this paper, we construct a neural network which consists of 60 samples of 3 key design variables, i.e. ${D_2}$,  ${b_2}$  and  ${\beta _2}$. 80\% of the samples are chosen as training set, 10\% as validation set, the last 10\% as testing set. Considering the scale of training data is small, the number of hidden neurons is determined to be 50, which is modest and can guarantee the overfitting is not happen. Bayesian Regularization is chosen as training algorithm. The result of fitting performance is shown in \ref{fig Training results of NN}, we can see the best training performance is 5.8994e-9 at epoch 1000, the gradient and error is close to zero, the $R$ is almost 1, which is satisfying for data fitting.
\begin{figure}[htb]
\centering
\subfigure[]{%
  \includegraphics[scale=0.22]{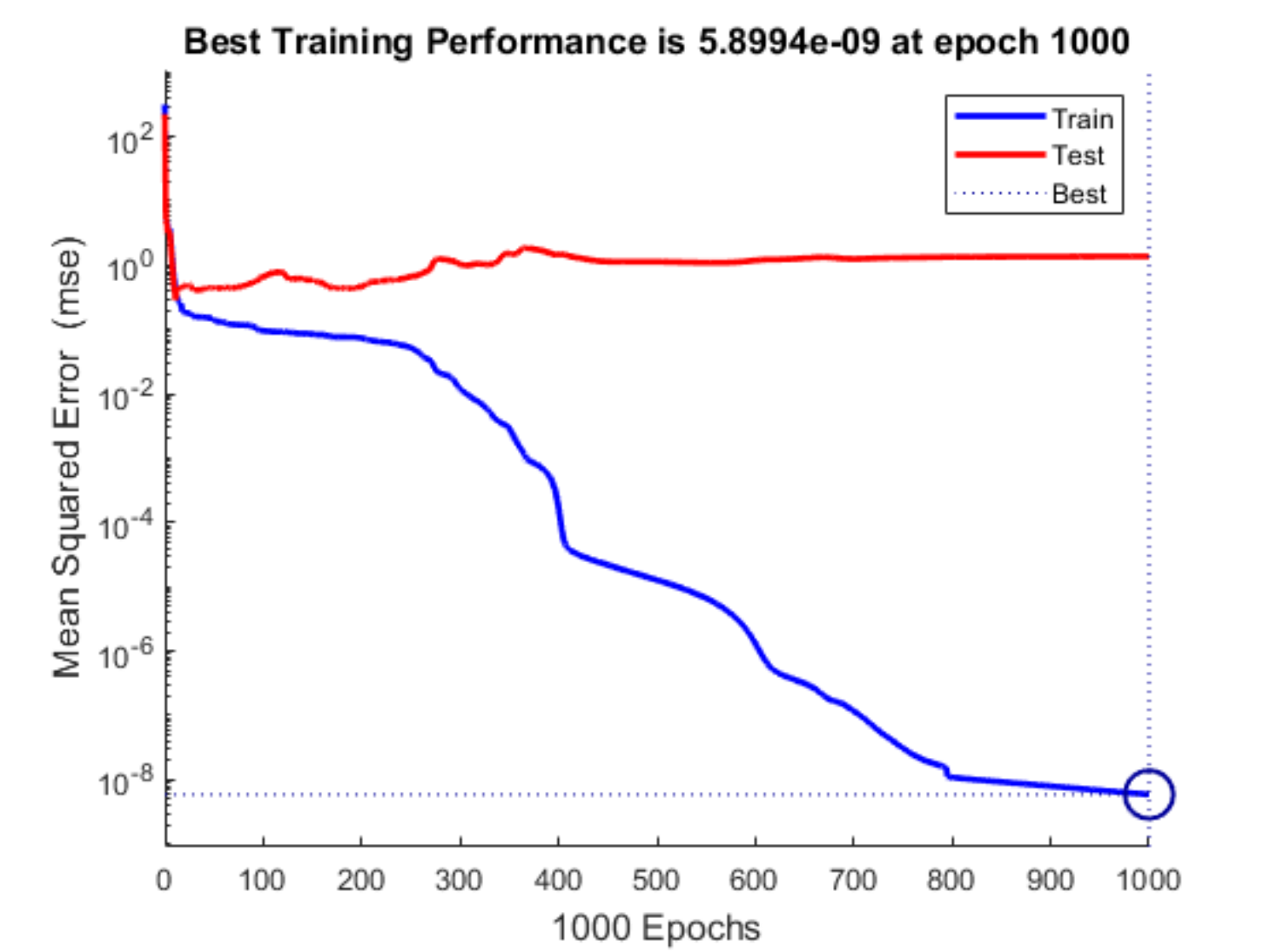}
  }
\quad
\subfigure[]{%
  \includegraphics[scale=0.22]{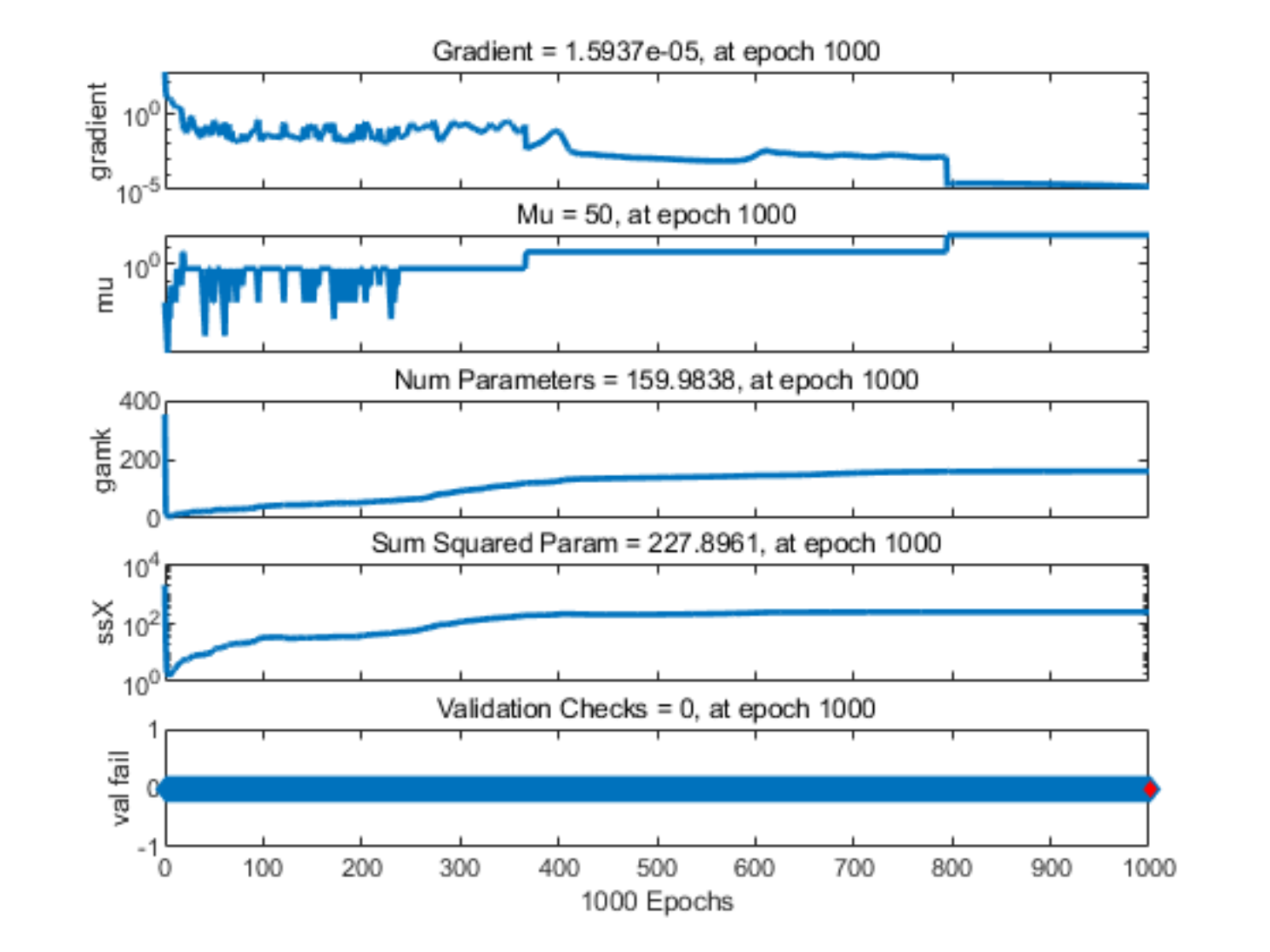}
  }

\subfigure[]{%
  \includegraphics[scale=0.22]{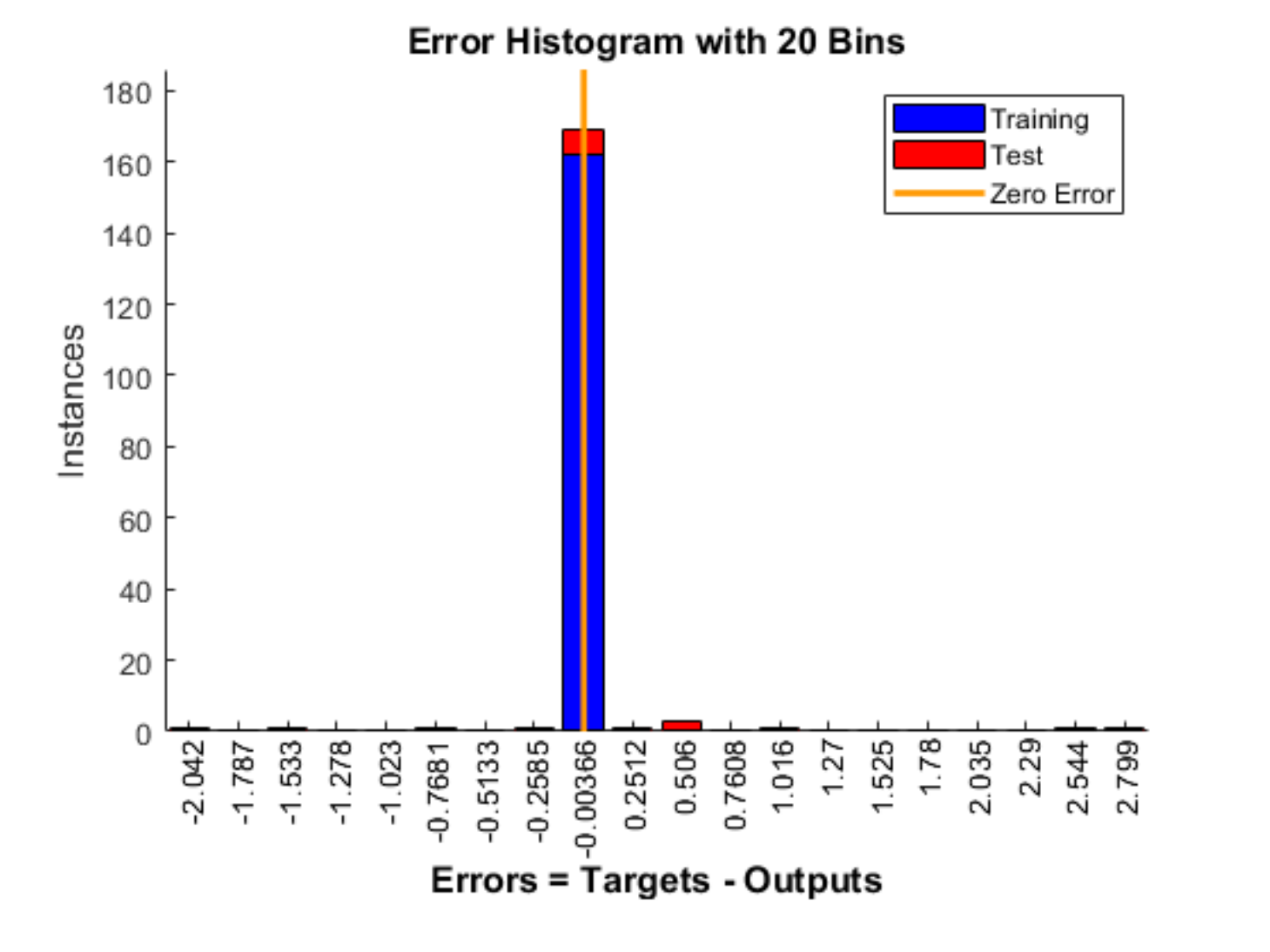}
  }
\quad
\subfigure[]{%
  \includegraphics[scale=0.18]{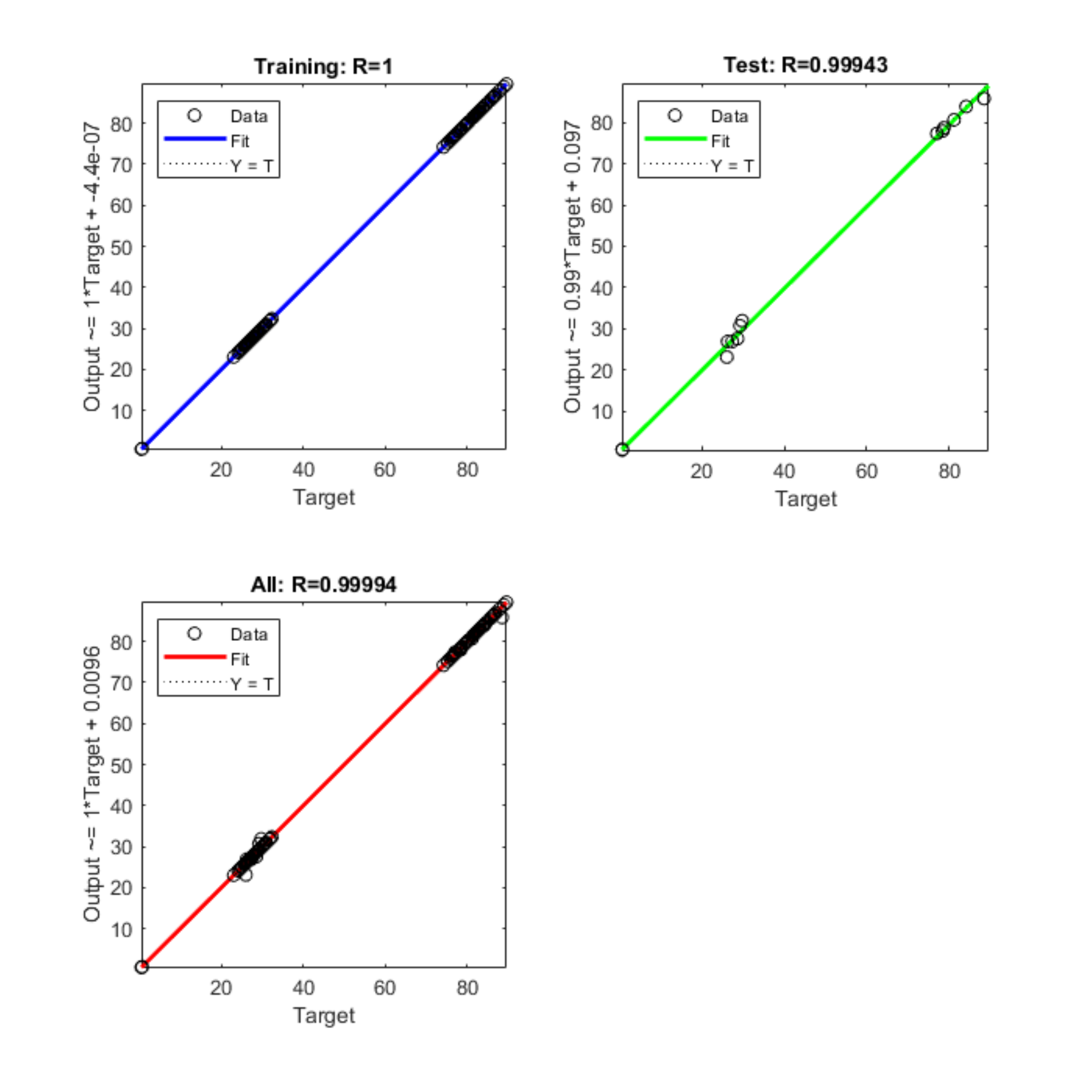}
  }
\caption{Training results of neural network. (a) Training Performance. (b) Training state. (c) Error histogram. (d) Regression.}
\label{fig Training results of NN}
\end{figure}
\section{Comparison of models for performance prediction}
\label{sec:3}
\subsection{Numerical simulation analysis of a sample centrifugal pump }

\begin{figure}[tbh]
	\centering
	\includegraphics[scale=1]{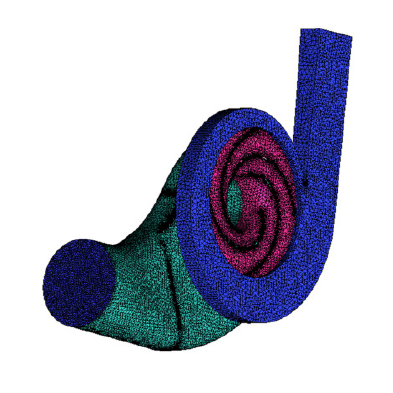}
	\caption{Comparison of different methods in terms of outlier detection accuracy.}
  \label{fig Gridmodel}
\end{figure}

The fluid domain of the first stage centrifugal pump is partitioned into suction chamber, impeller, volute and diffuser. The model of the sample pump is built and simulated by software CFturbo, ICEM and Fluent. The grid model is shown in Fig.~\ref{fig Gridmodel}. The RNG k-epsilon model is used for setting boundary conditions and simulation. The results of the grid independent verification is shown in the following Table~\ref{Table Grid independent}.

\begin{table*}[thp]\footnotesize
\centering
\caption{Grid independent verification} \label{Table Grid independent}
\addtolength{\tabcolsep}{4.8pt}
\begin{tabular*}{14.5cm}{ccccccc}
	\toprule[0.75pt]
	 Number         &       1      &  2         &       3       & 4       &    5    &      6          \\
   \midrule[0.5pt]
  Head ($m$)       &    81.54     &	 81.91    &   81.73      &	81.27   &	  81.56 &	  80.50      \\
	Efficiency (\%)  &    73.90     &	73.97	    &    74.44     &	74.85   &	74.74	  &   74.21      \\
	Time ($min$)     &   61         &	68        &	75           &	86      &	94      &	  99         \\
  \multirow{1}{*}{Performance value} &  76.7307 &	80.9646	 & 97.8905 &	116.7904 &	144.1980 &	170.7942 \\
	\bottomrule[0.75pt]
\end{tabular*}
\end{table*}

Looking at the calculation error and time, we realize that simulation calculation of the group NO.3 with the number of grids is the best. According to the result, the most effective meshing method is, the minimum grid size 0.5 mm, the maximum size of global grid 8 mm, the maximum grid size of blade 2 mm, and the maximum grid size of internal interface 4 mm. Besides, the grid quality is adjusted to assurance grid quality of 0.4 or more. The example training set is conducted numerical simulation utilizing DELL-R930 server for 40-cores parallel computing. The overall calculation time is about 40 hours.

Before comparing and testing the performance accuracy predicted by the above models, we implemented the numerical model validation based on another single stage centrifugal pump named HPB40. The design flow $Q = 10{m^3}/h$, Head $H = 45m$, rotational speed $n = 2900{\kern 1pt} {\kern 1pt} {\kern 1pt} r/min$, structure parameters of the pump are as follows:  ${D_2} = 190mm$,  ${b_2} = 6mm$,  ${\beta _2} = {28^ \circ }$,  ${D_h} = 28mm$,  ${D_s} = 50mm$,  ${b_1} \approx 16.5mm$,  ${\beta _1} = {30^ \circ }$,  $\varphi  = {150^ \circ }$,  ${D_3} = 1.10{D_2}$,  $Y = 4.5$,  ${b_3} = 17mm$. A numerical model of pump is established and simulated by CFD based on the above parameters. Therefore, the performance can be determined from the simulation results and verified with experiment data we have already gathered. The validation results are shown in Figure 6. It can be seen that the difference between the CFD and experiment is less than 3\% under the design flow and 6\% under all range of flow, so we can say the numerical model based CFD is quite accurate and fair. Therefore, the differences between CFD simulation results and prediction values of models are used to measure the prediction ability of different models in the following analysis.

\begin{figure}[htb]
\centering
\subfigure[]{%
  \includegraphics[scale=1.2]{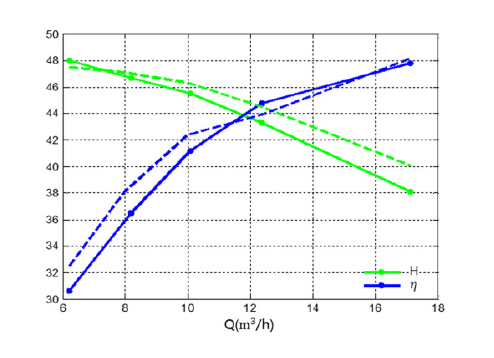}
  }
\quad
\subfigure[]{%
  \includegraphics[scale=1.2]{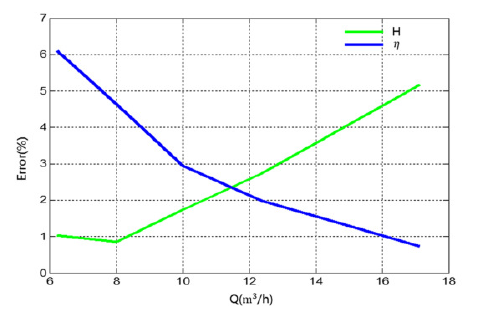}
  }
\caption{Validation of numerical model compared with experimental data. (a) Head(green) and efficiency(blue) comparison between CFD (dotted) and experiment (solid). (b) Head(green) and efficiency(blue) error of CFD relative to experiment.}
\label{fig Training results of NN}
\end{figure}

\subsection{Results and analysis}

\begin{figure}[tbh]
	\centering
	\includegraphics[scale=0.50]{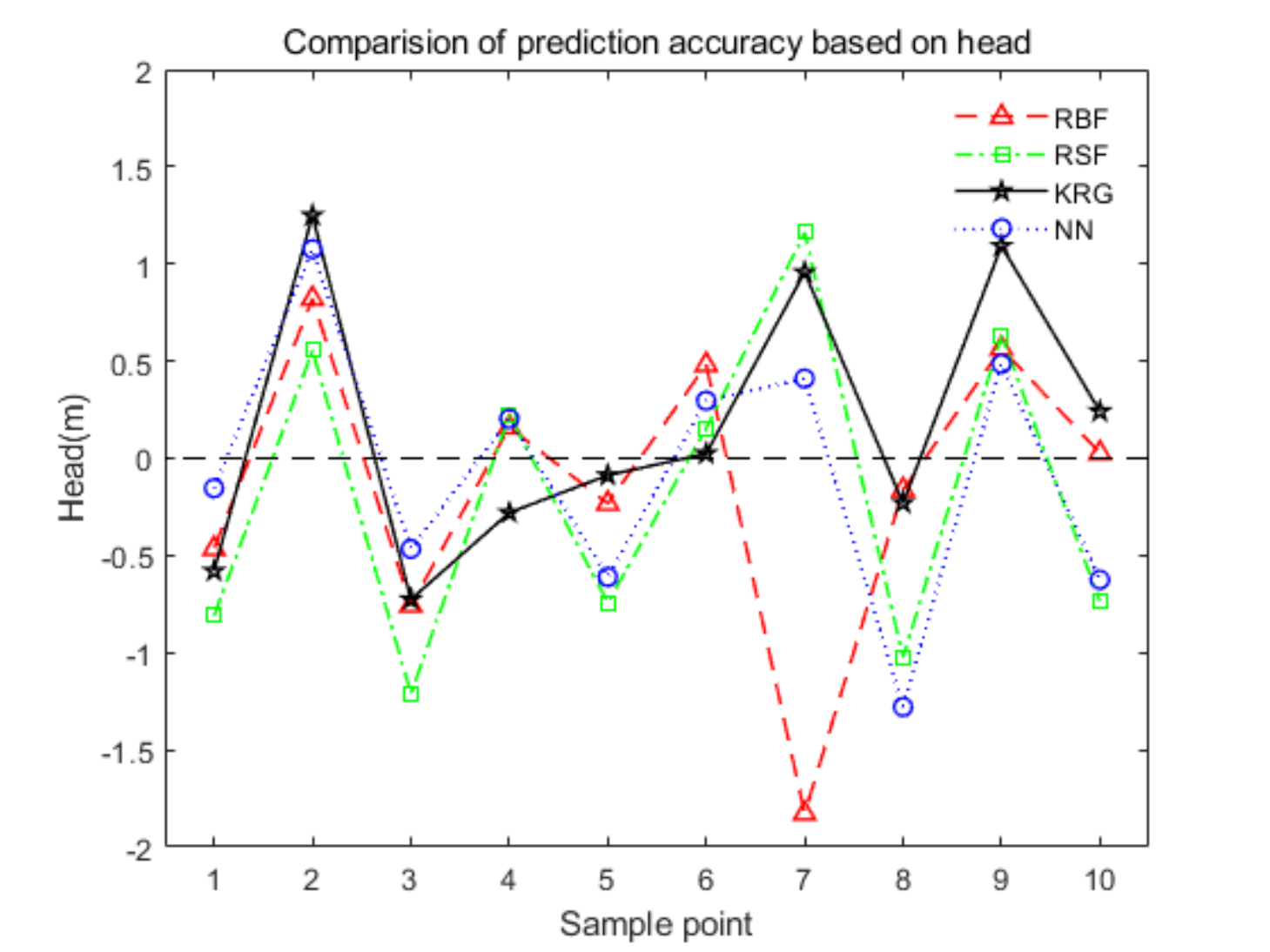}
	\caption{Comparison of prediction accuracy based on head.}
	\label{fig Comparison on head}
\end{figure}

\begin{figure}[tbh]
	\centering
	\includegraphics[scale=0.50]{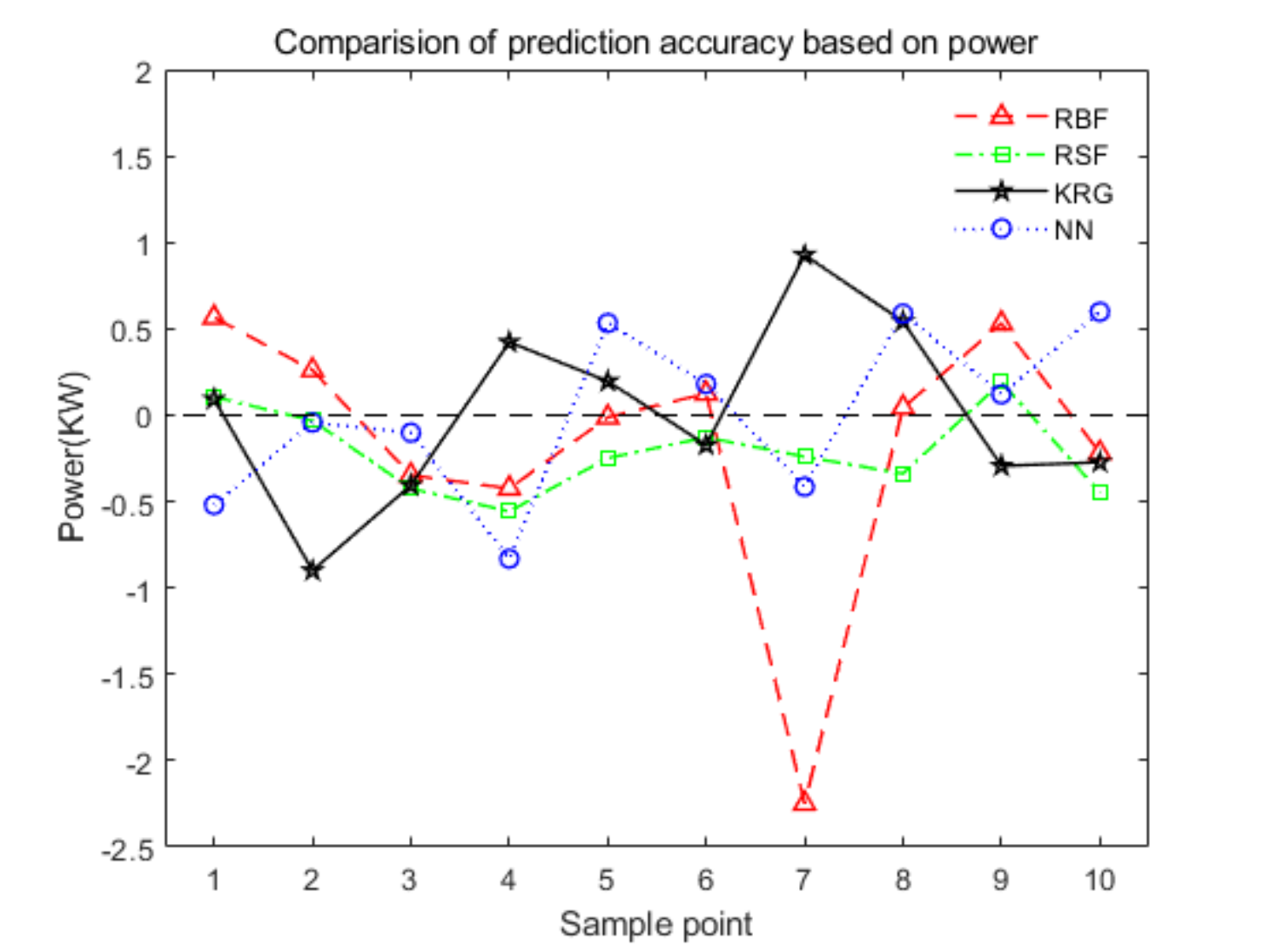}
	\caption{Comparison of prediction accuracy based on power.}
	\label{fig Comparison on power}
\end{figure}

For the propose that comparing and validating the prediction accuracy of the above three surrogate models and neural network, the Latin Hypercube sampling method is also used to randomly set up the test data for models in the design space. That takes another 10 samples of the key design variables, then numerical simulations are implemented to acquire the corresponding centrifugal pump performance value. The variable values of the test data are substituted into the RSF, RBF, KRG and NN models to obtain the predictions of the centrifugal pump performance and compared with the CFD numerical simulation. In order to compare the performance of models easily, the differences between predication values and CFD values of head and power  are plotted respectively. The results are shown in Fig.~\ref{fig Comparison on head}--\ref{fig Comparison on power}.


It shows that all four models can predict the head of pump well in Fig.~\ref{fig Comparison on head}, of which the NN model is best, the RSF has the maximum error. The RBF model has the overall minimum error except at sample point NO. 7, however, which has the lowest accuracy among all predictions. This problem can also be seen in Fig.~\ref{fig Comparison on power}, the RBF model also has the maximum error at point NO.7. Comparison of prediction accuracy based on power, the RSF model has the highest accuracy, NN is the second.

\begin{figure}[htb]
\centering
\subfigure[]{%
  \includegraphics[scale=0.5]{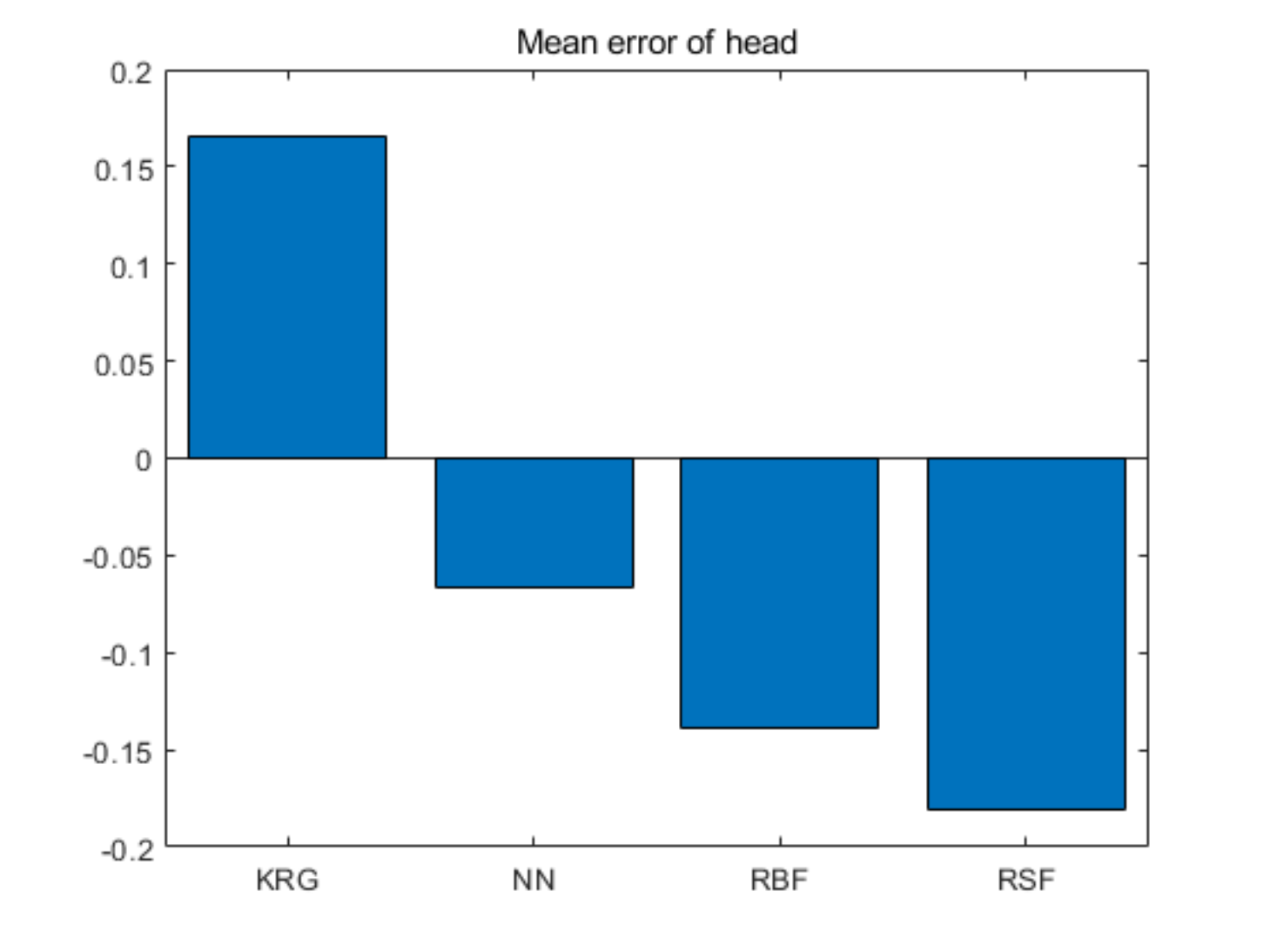}
  }
\quad
\subfigure[]{%
  \includegraphics[scale=0.5]{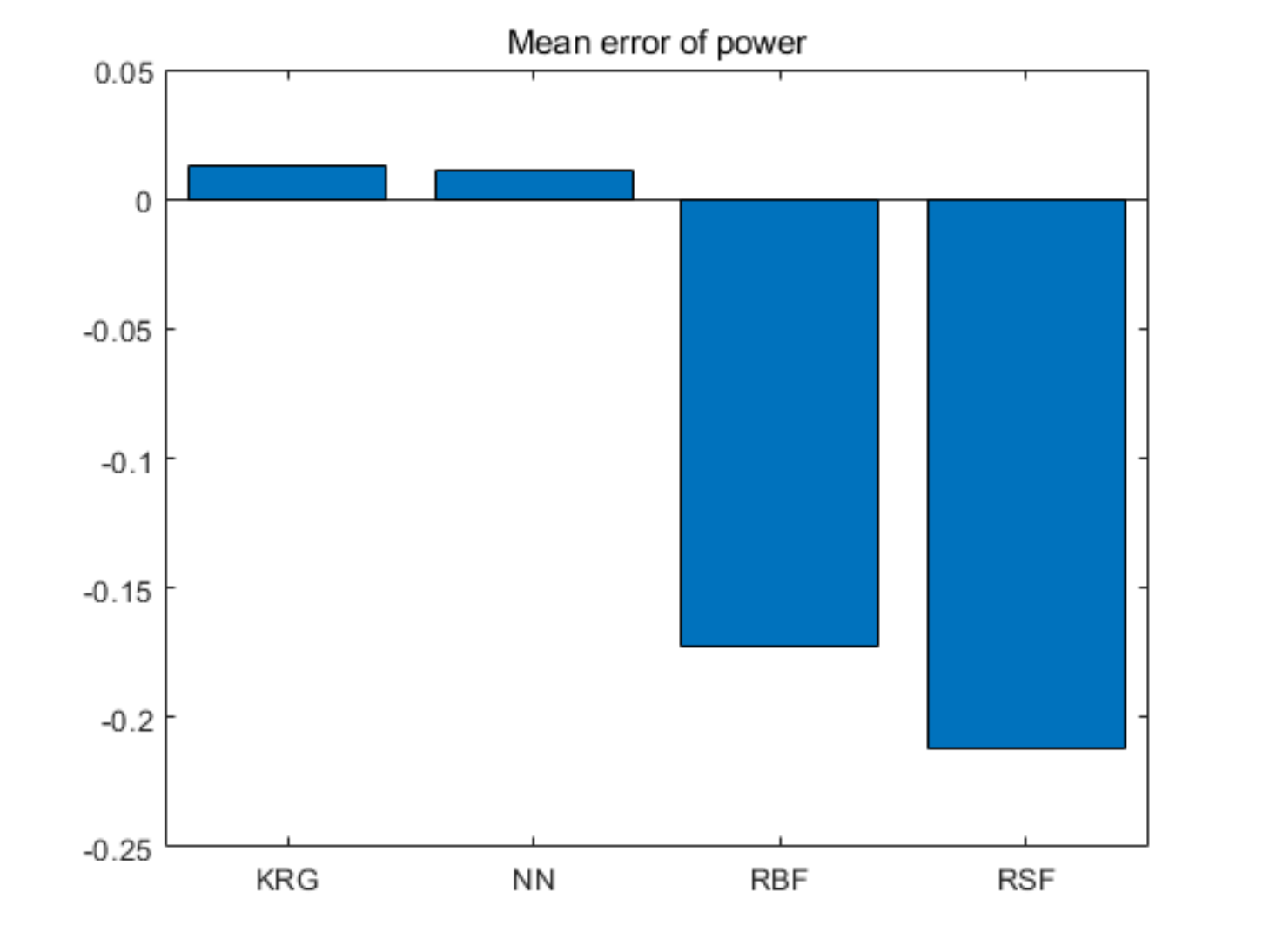}
  }
\caption{Mean error of models.  (a) Mean error based on head.  (b) Mean error based on power. }
\label{fig meanerror}
\end{figure}

In order to compare the models more specifically, the $RMSE$ and the $R^2$ of four models are shown in Table~\ref{table validity}. Comparing the root mean square error and the coefficient of determination of four models, and shown in above Fig.~\ref{fig Comparison on head}--\ref{fig Comparison on power}, the RSF and NN seems to be more accurate, RSF is the best prediction model. However, mean errors of four models are calculated for further analysis and shown in Fig.~\ref{fig meanerror}. Obviously, the NN model has the smaller mean error than RSF model in all predictions. It indicates that the prediction values of NN fluctuate in the smallest range near all kinds of performance values of CFD, i.e. head and power. Therefore, the NN model has better stability of predicting the high nonlinear relationship between input design variables and output prediction values of performance. In summary, the NN model has the best overall performance in multi-objective prediction of multistage centrifugal pump with the highest accuracy and stability.

\begin{table*}[thp]\footnotesize
\centering
\caption{Prediction models validity test. }
\label{table validity}
\addtolength{\tabcolsep}{4.8pt}
\begin{tabular*}{10cm}{cccccc}
	\toprule[0.75pt]
	\multirow{2}*{Prediction models}  & \multicolumn{2}{c}{$RMSE$ (\%)}  &  & \multicolumn{2}{c}{$R^2$ (\%)}   \\
	 \cmidrule[0.5pt]{2-3}  \cmidrule[0.5pt]{5-6}
	      & Head & Power  & &   Head & Power     \\
	\midrule[0.5pt]

	RBF		&  0.29  &  0.89 &  & 93.97  & 75.16  \\
    RSF		&  0.31  &	0.36 &  & 92.94  & 95.56  \\
    KRG		&  0.27  &	0.58 &  & 94.78  & 89.63  \\
    NN		&  0.26  &	0.53 &  & 95.21  & 91.11  \\
	\bottomrule[0.75pt]
\end{tabular*}
\end{table*}
\section{neural network model with data augmentation}
\label{sec:4}
From above analysis it is clear that neural network is a good model for fitting the high nonlinear connections between design variables and multi-objective performance values. Moreover, with the rapid development of AI, more and more intelligent algorithms are proposed and neural network evolves continuously in recent years. Therefore, neural network model is a promising model and supposed to become more accurate. For the propose of training a model, a fair amount of data is undoubtedly needed. In big data problems, when facing the huge number of data, such as natural language processing (NLP) and image recognition (IR), training cost should be taken into consideration. In contrast, insufficiency of data is a common dilemma in optimization problem of mechanical design. There are plenty of combinations of different design variables in design range, but the simulation is resource-consuming, so the common resolution is Latin hypercube sampling method to randomly set up the test data for models in the design space. The problem of lacking data isn't solved unless more simulations are implemented. Aiming at this problem, a neural network model with data augmentation is proposed by introducing the interpolation method considering the system error.

\begin{figure}[tbh]
	\centering
	\includegraphics[scale=0.5]{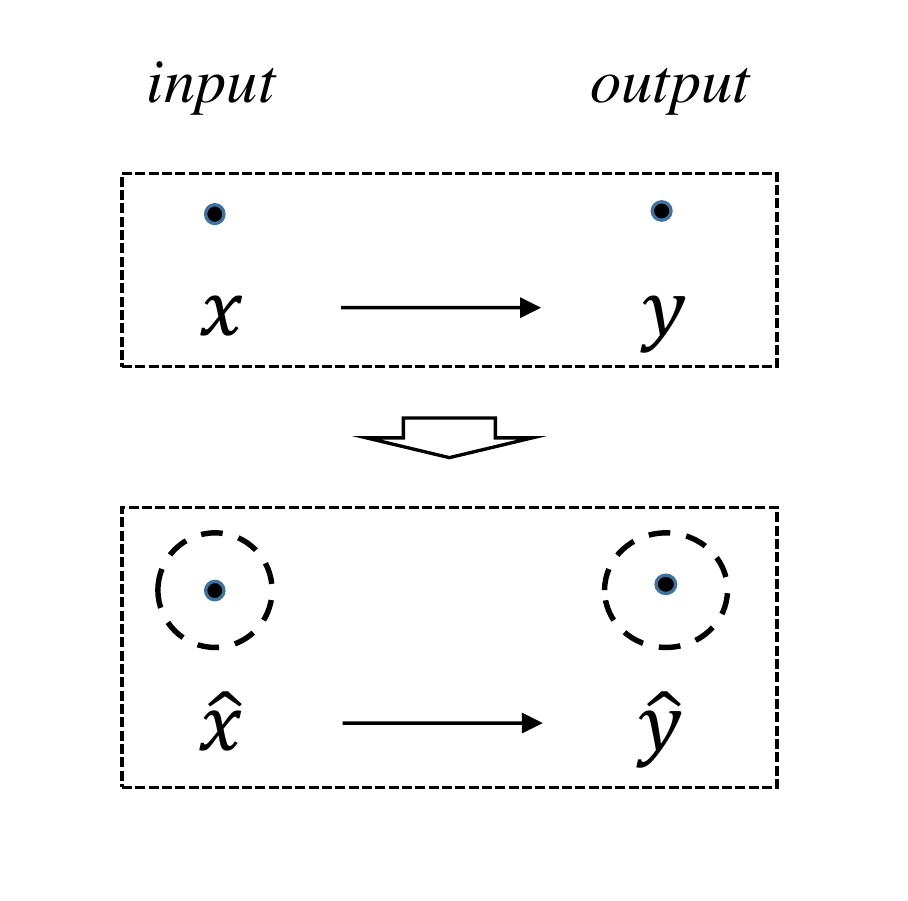}
	\caption{Schematic diagram of NNDA.}
	\label{fig schematic of NNDA}
\end{figure}

The inspiration of data augmentation arises from the error analysis of the simulation processing. In the light of engineering experience of simulation, there are many reasons lead to the system error, such as truncation error caused by using discrete equations with different numerical methods instead of original equations, rounding error caused by insufficient storage accuracy of computer and so on. In the specific engineering problems of simulation, we believe that the error rate is acceptable when it is less than 5\%. Besides, the performance value base on simulation result is adopted to training the model. Therefore, the extremely accurate model whose error is very close to zero on training set can't perform very well on the generalized set. On the basis of above analysis, we can conclude that the simulation result is an approximate response for given input design variables $x$. As a result, to get more useful information for training model, we should give heed to the distributions of input and output, which are neglected in previous studies. The schematic is shown in Fig.~\ref{fig schematic of NNDA}. A reasonable error in a small range is introduced at each sample point of each attribute for input and output. Consequently, it intensifies the high nonlinear connections between input and output.

\begin{algorithm}\small
\centering
\setstretch{1.5}
\caption{Algorithm for data augmentation} \label{alg dataaug}
\begin{algorithmic}[1]
	\STATE {$IF \gets \text{0.025}$}
	\algorithmiccomment{determine interpolation factor}
	\LOOP
    \STATE{ $att \gets  \text{attribute in input $x$ and output $y$}$}
      \FOR{$P \gets \text{point to be processed of one attribute}$}
        \STATE {calculate the difference between P and other points}\label{stepdif}
        \STATE {$inter \gets \text{the minimum of difference in step.\ref{stepdif}} $}
		    \STATE{$bias \gets \text{$inter*IF$} $} \algorithmiccomment{determine interpolation internal}
        \STATE{$newp1 \gets  \text{$P+bias$}$}
        \STATE{$newp2 \gets  \text{$P-bias$}$} \algorithmiccomment{get new point}
        \STATE{$newdata \gets  \text{$[newp1,newp2]$}$} \algorithmiccomment{save}
		\ENDFOR
		\STATE{do line~4--11 for all attribute, obtaining augmented data of all attribute}
	\ENDLOOP
	\STATE {save the augmented data for training}
\end{algorithmic}
\end{algorithm}

\begin{figure}[htb]
\centering
\subfigure[]{%
  \includegraphics[scale=0.22]{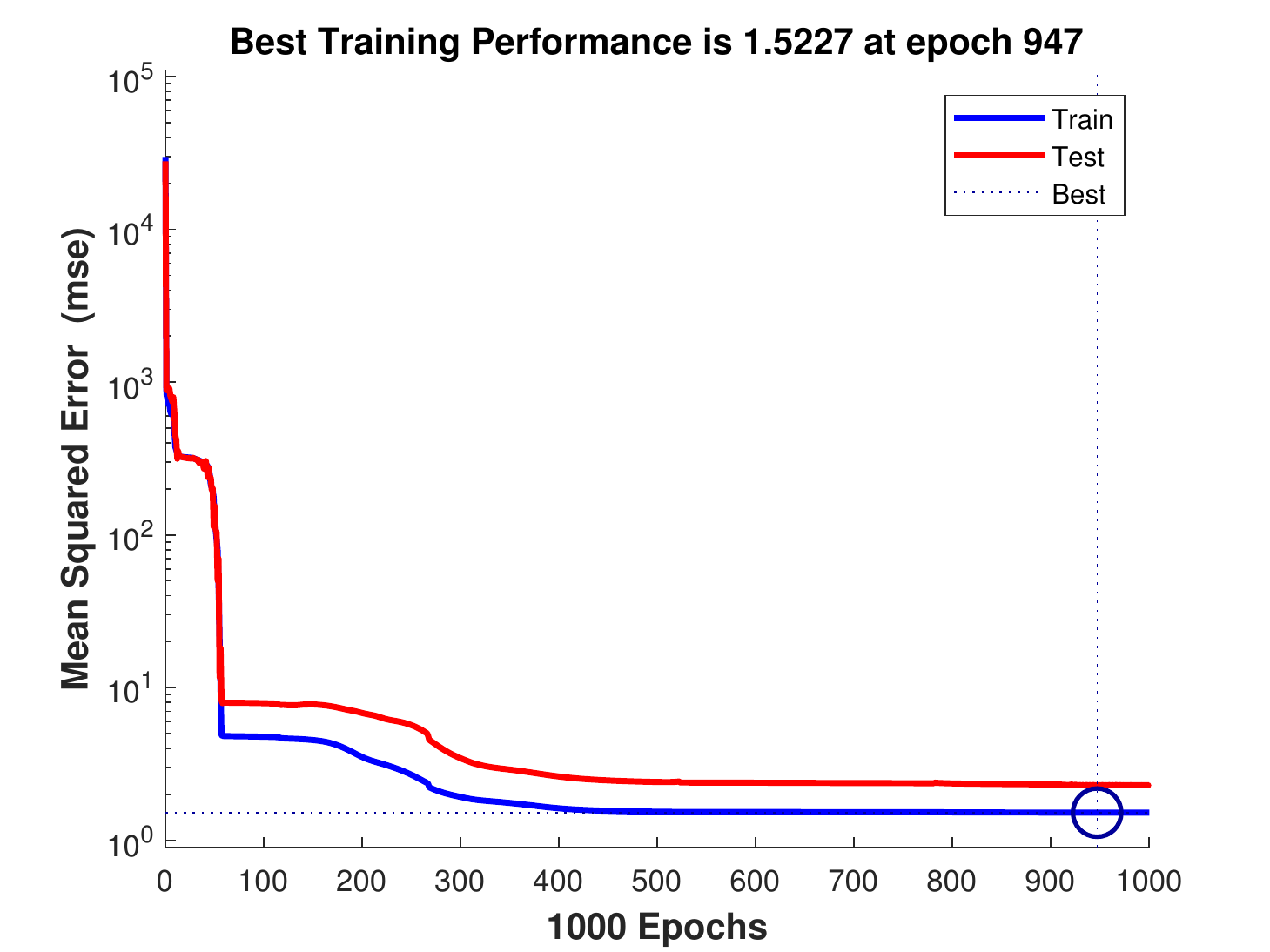}
  }
\quad
\subfigure[]{%
  \includegraphics[scale=0.22]{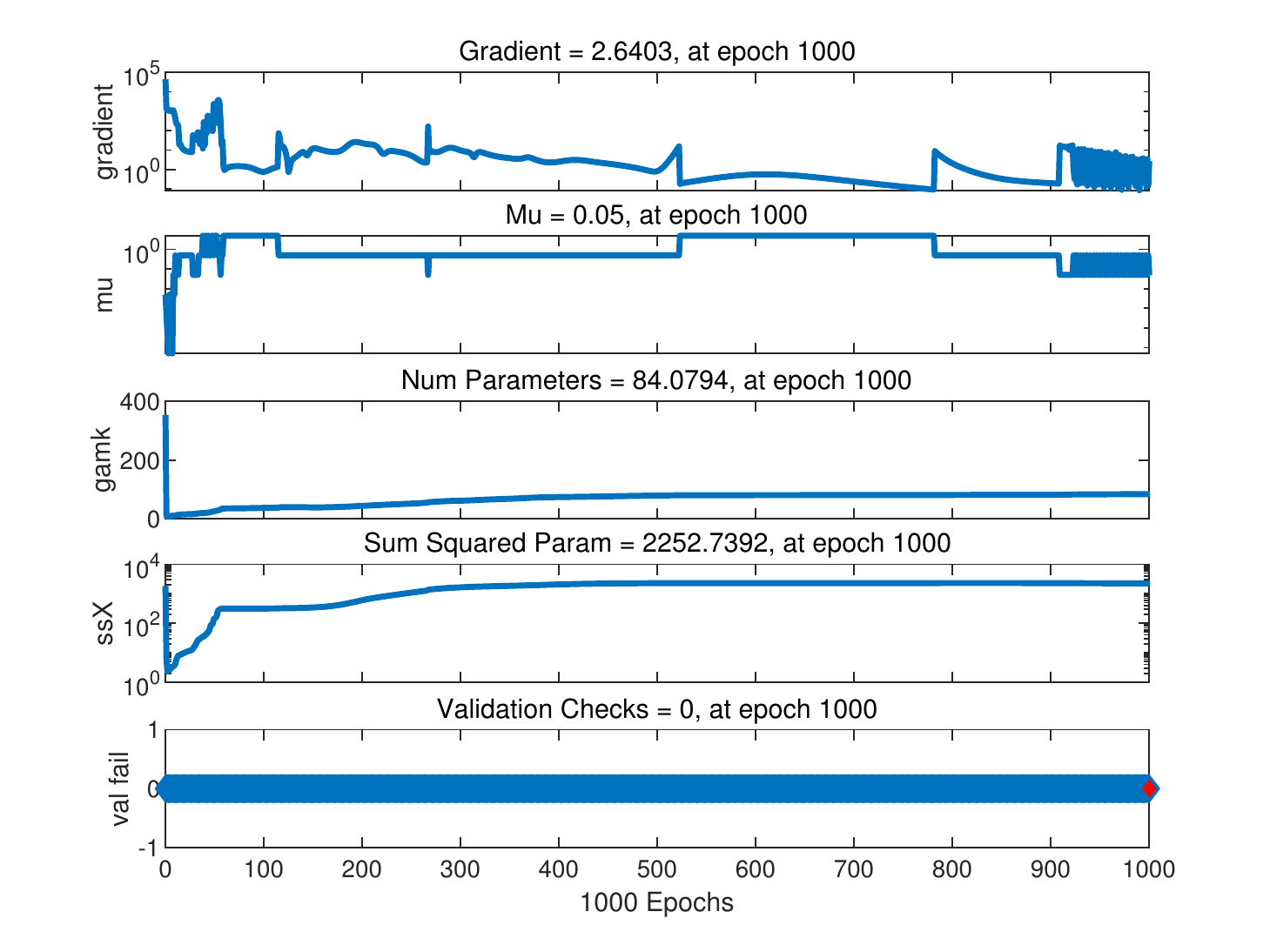}
  }

\subfigure[]{%
  \includegraphics[scale=0.22]{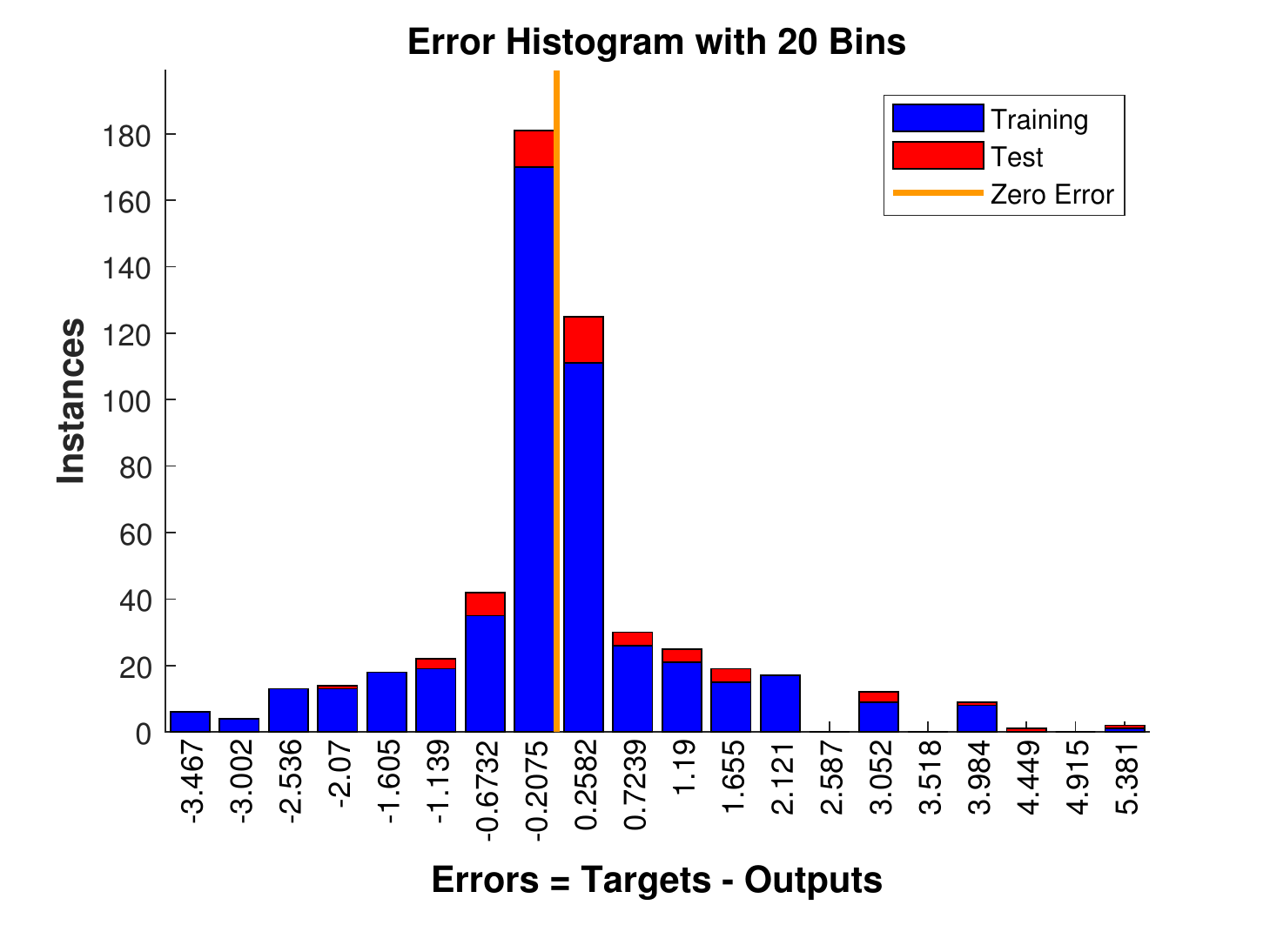}
  }
\quad
\subfigure[]{%
  \includegraphics[scale=0.18]{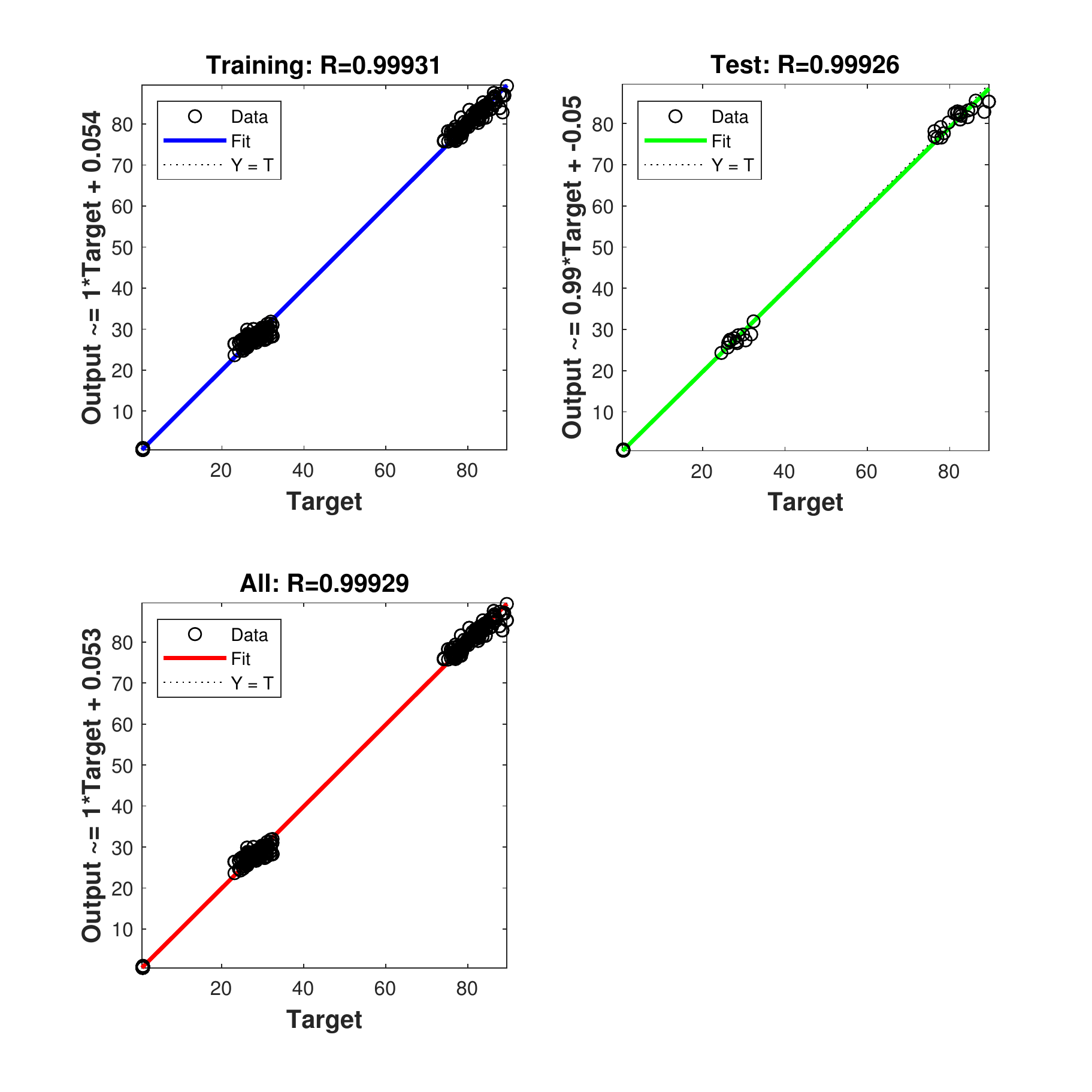}
  }
\caption{Training results of neural network with data augmentation. (a) Training Performance. (b) Training state. (c) Error histogram. (d) Regression.}
\label{fig Training results of NNDA}
\end{figure}

The brief introduction of data augmentation is shown in Algorithm~\ref{alg dataaug}. To begin with, the interpolation factor is defined as 0.025, i,e. 2.5\%. For the reason that we want the point values to fluctuate within the range of 5\%, as mentioned above, $\pm~2.5\%$ is logical. Then the different attributes of input $x$ (design variables) and output $y$ (simulation results) is to be operated separately. For each value in attributes, the different betweent it and other points is calculated and the minimum value is choosed. This step takes the distribution scale of each attribute into consideration and garuntees the introduced error is restrained. Next the minimum value is multiplied by the interpolation factor so the bias is obtained. The two new data point are acquired by the point P plus and minus the bias. Finally we loop the above steps and get another 120 samples. By this algorithm, the model triple the training data and train the neural network under the same settings in Section~\ref{section construction of models}. The result is shown in fig.~\ref{fig Training results of NNDA}.

\begin{figure}[tbh]
  \centering
  \includegraphics[scale=0.50]{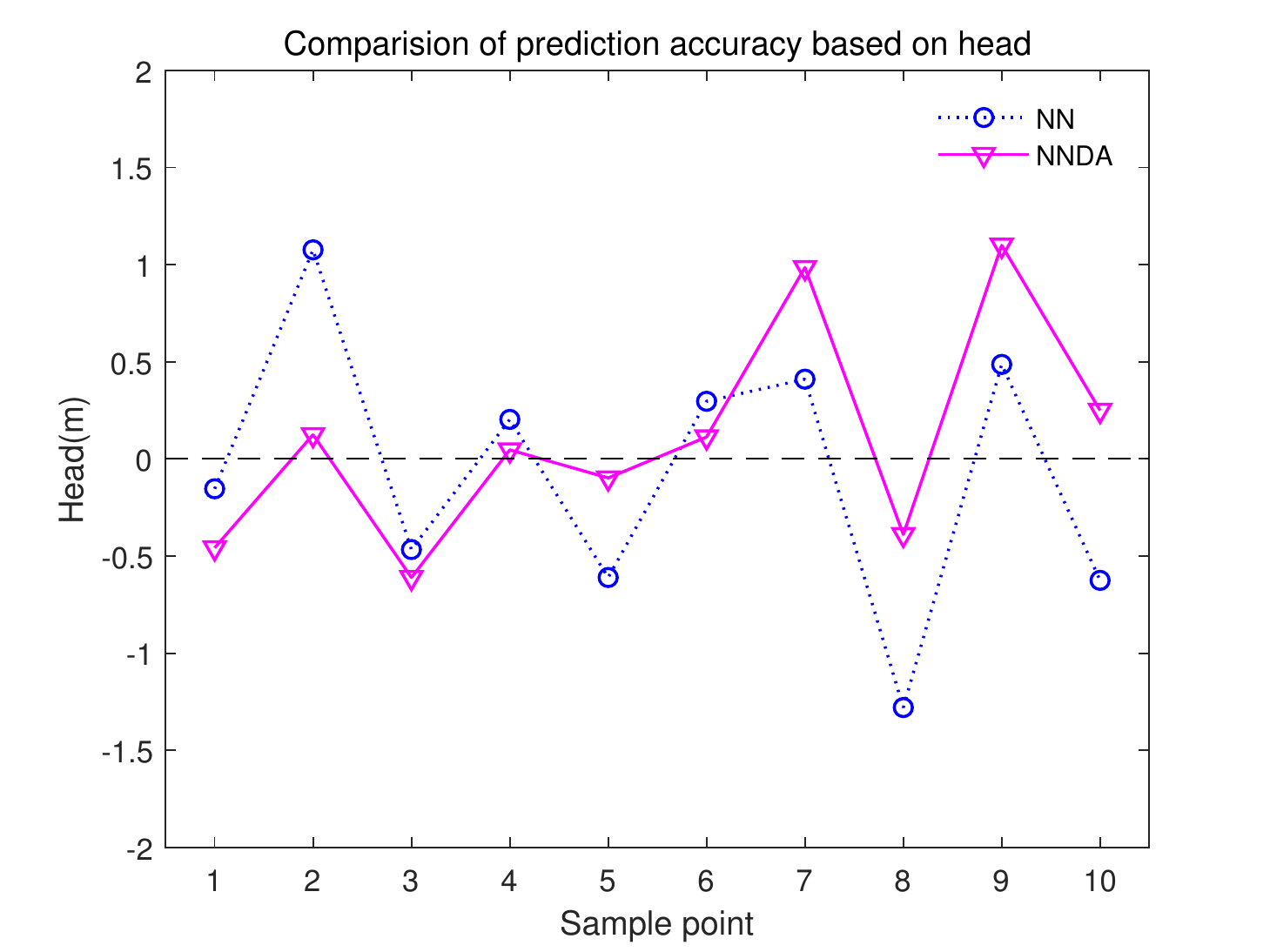}
  \caption{Comparison of prediction accuracy based on head.}
	\label{fig headnnnnda}
\end{figure}

\begin{figure}[tbh]
	\centering
	\includegraphics[scale=0.50]{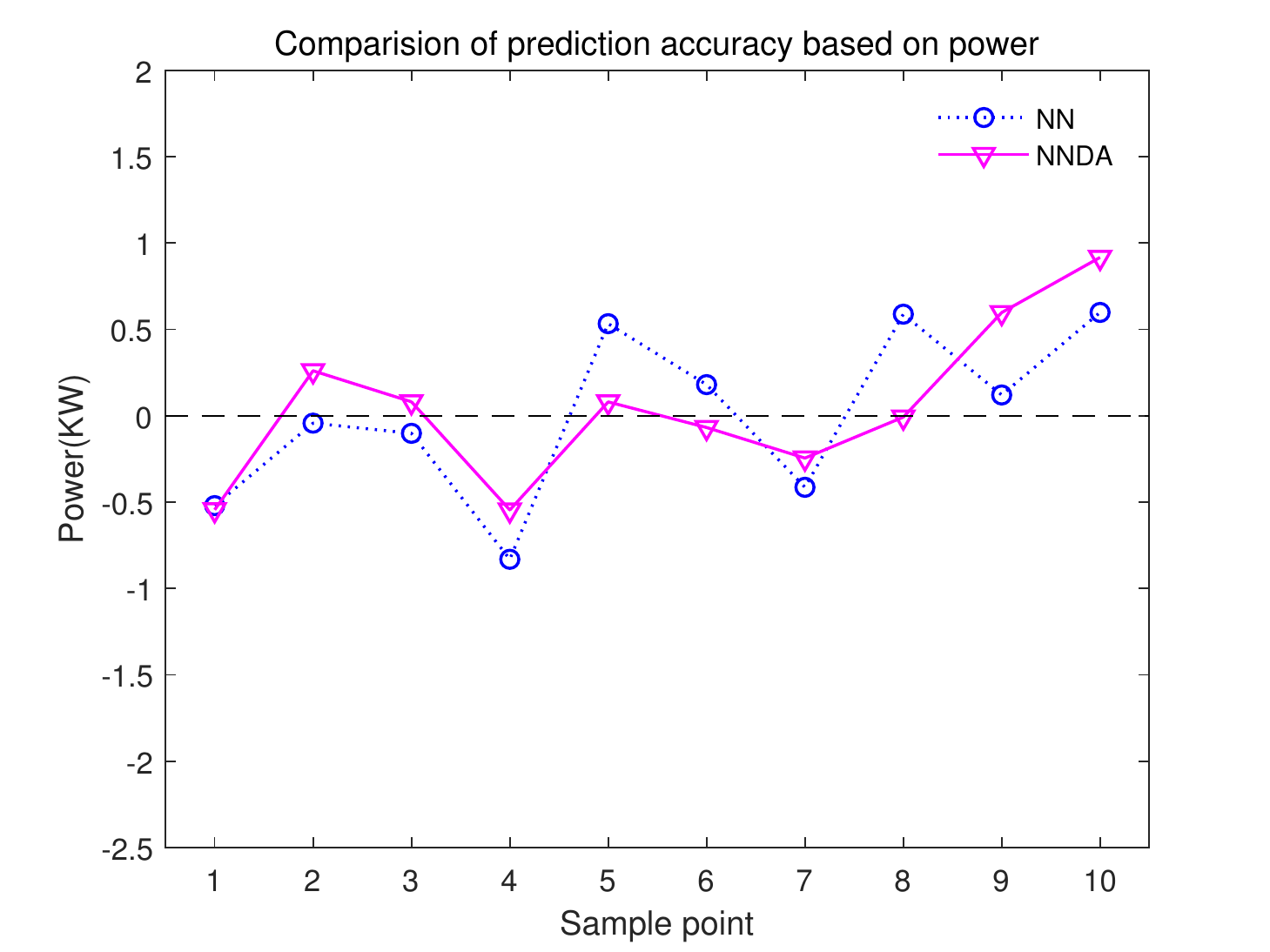}
	\caption{Comparison of prediction accuracy based on power.}
	\label{fig powernnnnda}
\end{figure}

From Fig.~\ref{fig Training results of NN} and Fig.~\ref{fig Training results of NNDA} it shows that the performance of NNDA is better than NN in test set. However, the error histogram shows that the NNDA has greater error than NN, this is understandable because we intruduced the error of 5\% at each sample point. The performance comparison on another 10 test set of NNDA and NN is shown in Fig.~\ref{fig headnnnnda}--\ref{fig powernnnnda}.

\begin{table*}[thp]\footnotesize
\centering
\caption{Prediction models validity test. }
\label{table validity}
\addtolength{\tabcolsep}{4.8pt}
\begin{tabular*}{10cm}{cccccc}
	\toprule[0.75pt]
	\multirow{2}*{Prediction models}  & \multicolumn{2}{c}{$RMSE$ (\%)}  &  & \multicolumn{2}{c}{$R^2$ (\%)}   \\
	 \cmidrule[0.5pt]{2-3}  \cmidrule[0.5pt]{5-6}
	      & Head & Power  & &   Head & Power     \\
	\midrule[0.5pt]
    NN		  &  0.26  &	0.53 &  & 95.21  & 91.11  \\
    NNDA		&  0.21  &	0.50 &  & 96.67  & 92.10  \\
	\bottomrule[0.75pt]
\end{tabular*}
\end{table*}

It can be seen that the NNDA performs a little better than NN from Fig.~\ref{fig headnnnnda}--\ref{fig powernnnnda}, and the $RMSE$ and $R^2$ are calculated for further analysis, shown in Table~\ref{table validity}. It is clearly that NNDA model works better in both head and power predictions, though the performance of NN is good enough in previous comparison.
This result proves that the data augmentation method is reasonable and effective. Moreover, how to continuously optimize the algorithm of NNDA is still a meaningful direction in future research.

\section{Conclusion}
\label{sec: conclusion}
This paper focus on the multi-objective hydraulic performance predictions of a multistage centrifugal pump based on surrogate models and neural network. In order to determine key design variables, sensitivity analysis of design variables was performed based on the hydraulic loss model. The key design variables $x = \left[ {{D_2},{b_2},{\beta _2}} \right]$  were determined and the prediction values are head and power. For the propose of better fitting the high nonlinear relationships between input x and output y, the three surrogate models, the NN model NN model (neural network model)  is constructed comparing with the RBF (radial basis Gaussian surface model), the RSF (quadratic response surface model) and the KRG (Kriging model). Comparing the prediction accuracy based on head, it shows that all four models can predict the head of pump well, of which the NN model is the best, the RSF has the maximum error. The RBF model has the overall minimum error except at sample point NO. 7, however, which has the lowest accuracy. Comparison of prediction accuracy based on power, the RBF model also has the maximum error at sample point NO.7, the RSF model has the highest accuracy, NN is the second.

Comparing the root mean square error and the coefficient of determination of four models, the RSF and NN are more accurate, RSF is the best prediction model. Mean errors of four models are calculated and the results shows that the NN model has the smaller mean error than RSF model in all predictions. It indicates that the prediction values of NN fluctuate in the smallest range near all kinds of performance values of CFD, i.e. head and power. Therefore, the NN model has better stability of predicting the high nonlinear relationship between input design variables and output prediction values of performance. In conclusion, the NN model has the best overall performance in multi-objective prediction of multistage centrifugal pump with the highest accuracy and stability.

Aiming at the problem of lacking data in engineering practice, a neural network model with data augmentation is proposed by introducing the interpolation method considering the system error. According the engineering experience we takes the distribution scale of each attribute into consideration, allow point values to fluctuate and restrain the errors within the range of 5\%. For each value in attributes of input $x$ (design variables) and output $y$ (simulation results), the differences between it and other points are calculated. The minimum value is choosed and multiplied by the interpolation factor so the bias is obtained. The two new data point are acquired by the point plus and minus the bias. By this algorithm, the model can triple the training data.

Comparing with the former NN model, NNDA model works better in both head and power predictions, though the performance of NN is good enough in previous comparison. The preformance of NNDA model becomes more accuracy in term of the $RMSE$ and $R^2$ index values. In summary, it proves that the data augmentation is an effective method, which helps training model get triple data and improve the prediction accuracy without more simulation costs.

Building and fitting the model is the key step for optimization of multistage centrifugal pump multi-objective prediction. As a powerful and promising tool to fit the highly nonlinear relationship between the key design variables and the pump external characteristic values, neural network evolves rapidly and its prediction performance is expected to improve continuously. It can be more accurate and powerful with data augmentation method, which can apply to optimization problems of multistage pump and generalize to other finite element analysis optimization problems. Moreover, continuously optimize the algorithm of NNDA is still a meaningful direction in future research.

\bibliographystyle{unsrt}  

\end{document}